\documentclass[12pt]{article}
\usepackage{amsfonts, amsmath, amssymb, amsthm}
\usepackage{color}
\usepackage{datetime}
\usepackage{dsfont}
\usepackage{graphicx}
\usepackage{scrtime}
\usepackage[T1]{fontenc}

\newcommand{\R}{\mathbb{R}}  

\newcommand{\bfr}{{\bf r}}

\newcommand{\bfu}{{\bf u}}

\newcommand{\bfR}{{\bf R}}

\newcommand{\bra}[1]{\ensuremath{\langle #1 \vert}}
\newcommand{\ket}[1]{\ensuremath{\vert #1  \rangle}}

\begin{document}
\title{On connecting density functional approximations to theory} 
\author{Andreas Savin \\ Laboratoire de Chimie Th\'eorique \\ CNRS and Sorbonne University \\ 4 place Jussieu, F-75252 Paris, France \\
 andreas.savin@lct.jussieu.fr }
\date{\currenttime \; \today}

\maketitle
 
\begin{small}
 submitted for publication in: \\
{\em Density Functional Theory} \\
 Eds. Eric Canc\'es, Lin Lin, Jianfeng Liu, and Gero Friesecke \\
 Springer series on Molecular Modeling and Simulation, Vol. 1
\end{small}

\begin{abstract}

Usually, density functional models are considered approximations to density functional theory,
However, there is no systematic connection between the two, and this can make us doubt about a linkage.
This attitude can be further enforced by the vagueness of the argumentation for using spin densities.
Questioning the foundations of density functional models leads to a search for alternative explanations.
Seeing them as using models for pair densities is one of them. Another is considering density functional approximations 
as a way to extrapolate results obtained in a model system to those of a corresponding physical one.
 
\end{abstract}

\begin{flushright}
 Dedicated to Jean-Paul Malrieu \\
 on his 80th birthday
\end{flushright}

\section{Introduction}
\subsection*{}
\subsubsection*{On approximations in DFT}
Density functional theory is here.
It has changed the way the computation of electronic systems is seen by the scientific community.
It has a sound theoretical foundation.
However, following exact theory is more complicated than solving the Schr\"odinger equation. 
Furthermore, it does not tell us how to produce systematically approximations.
Usual approximations are convenient and (to a large degree) successful, but how to improve them?

\subsubsection*{Excuses}
This is not a review. 
References are erratic and biased.
Own publications dominate, not because they are more important, but because they are only given to complement argumentation. 

\subsubsection*{Summary}
After giving the notations, and reminding the Hohenberg-Kohn theorem, some practical solutions are recalled, such as the 
 decomposition of the universal functional, in order to comply with different physical requirements. 
It is argued that this does not necessarily solve the problem.
Refinements, such as using the spin density as a supplementary variable, are discussed.
It is argued that the need for these refinements may hide a different foundation for the approximations.
In order to introduce a ``systematic'' way to approach the physical Hamiltonian, model Hamiltonian are defined that via an adjustable 
 parameter approach the physical Hamiltonian.
Finally, examples show that simple mathematical recipes provide a quality similar to that of density functional approximations.

\section{Schr\"odinger equation and notations}

We start with a Schr\"odinger equation:
\begin{equation}
 H \Psi = E \Psi
\label{eq:schroedinger}
\end{equation}
The wave function $\Psi$ depends on the coordinates of the electrons $\bfr_1, \bfr_2, \dots, \bfr_N$ and their spins.
We will be mainly concerned with ground state eigenvalues, $E=E_0$.
We consider Hamiltonians of the form 
\begin{equation}
 H = T + V + W
\label{eq:hamiltonian}
\end{equation}
$T$ is the operator for the kinetic energy,
\begin{equation}
 T = -\frac{1}{2} \sum_{i=1}^N \nabla_i^2
\label{eq:t}
\end{equation}
$V$ is a local one-particle potential
\begin{eqnarray}
  V & = & \sum_{i=1}^N v(\bfr_i) \\
    & = & \int_{\R^3} v(\bfr) \hat{\rho}(\bfr) \, d\bfr 
\label{eq:1potential}
\end{eqnarray}
The density operator, $\hat{\rho}(\bfr)$, can be written using Dirac's $\delta$ function,
\begin{equation}
 \hat{\rho}(\bfr) = \sum_{i=1}^N \delta(\bfr - \bfr_i)
\end{equation}
its expectation value is the density
\begin{equation}
 \rho(\bfr) = \bra \Psi \hat{\rho}(\bfr) \ket \Psi 
 \label{eq:rho}
\end{equation}
Please notice that it integrates to $N$,
\begin{equation}
 N = \int_{\R} \rho(\bfr) \, d\bfr
 \label{eq:Nfromrho}
\end{equation}
$W$ is a two-particle local potential,
\begin{eqnarray}
 W & = & \sum_{i<j}^N w(\vert \bfr_i -\bfr_j \vert) \\
   & = & \frac{1}{2} \int_{\R^3} \int_{\R^3}  \; w(\vert \bfr_i -\bfr_j \vert) \hat{P}(\bfr,\bfr') \,  d\bfr \, d\bfr'
\label{eq:2potential}
\end{eqnarray}
where $\hat{P}$ is the pair density operator,
\begin{equation}
  \hat{P}(\bfr,\bfr')  =   \sum_{i \ne j}^N \delta(\bfr - \bfr_i) \delta(\bfr' - \bfr_j) 
   \label{eq:pop}
\end{equation}
The pair density is
\begin{equation}
 P(\bfr,\bfr')  =  \bra{\Psi} \hat{P}(\bfr,\bfr') \ket{\Psi}
 \label{eq:p}
\end{equation}
As the interaction depends only on the distance between particles, often the dependence of $P$ on $\bfr$ is reduced to that on 
 $u=\vert \bfr -\bfr' \vert$, using instead of $\hat{P}$ the spherically averaged operator
\begin{equation}
 \hat{P}_{sph}(\bfr,u)  =   \sum_{i \ne j}^N \delta(\bfr - \bfr_i) \delta(\vert \bfr -\bfr' \vert-u)
 \label{eq:psphop}
\end{equation}
yielding
\begin{equation}
 P_{sph}(\bfr,u)  =  \bra{\Psi} \hat{P}_{sphe}(\bfr,u) \ket{\Psi}
 \label{eq:psph}
\end{equation}
Going one step further, one can also integrate over $\bfr$, to obtain the system-average
\begin{equation}
 P_{sys}(u)  =  \int_{\R^3} \bra{\Psi} \hat{P}_{sphe}(\bfr,u) \ket{\Psi} \, d\bfr
 \label{eq:psys}
\end{equation}


For the electronic systems, $V=V_{ne}$, or $v=v_{ne}$, describes the Coulomb interaction between the nuclei and the electrons,
\begin{equation}
 v_{ne}(\bfr) = - \sum_A \frac{Z_A}{\vert {\bf R}_A - \bfr \vert}
\label{eq:vne}
\end{equation}
$A$ is an index for the nuclei, $Z_A$ their nuclear charge, and ${\bf R}_A$ their position.
Also, $W=V_{ee}$, or $w=v_{ee}$, describes the Coulomb interaction between electrons
\begin{equation}
 v_{ee}(\vert \bfr_i - \bfr_j \vert) = \frac{1}{\vert \bfr_i - \bfr_j \vert}
\label{eq:vee}
\end{equation}

To characterize a given electronic system, one has only to specify $N$ and $v_{ne}$. 


Model systems are considered below, where $v \ne v_{ne}$ and $w \ne v_{ee}$.
Of course, in this case the energies and wave functions depend also on the choice of $v$ and $w$. 
No change of the non-local one-particle operator $T$ is considered in this chapter, 
 but such modifications can be found in the literature (see, e.g., \cite{GutSav-PRA-07} for a density functional context).
The ground state energy can be also obtained using the variational principle,
\begin{equation}
  E[v,w,N] = \min_\Psi \bra \Psi H  \ket \Psi = \min_\Psi \bra \Psi T + V + W  \ket \Psi 
\label{eq:e0-var}
\end{equation}

\section{The density functional viewpoint}

\subsection{Hohenberg-Kohn Theorem}

In order to see how density functional theory can be useful, one generally argues using the Hohenberg-Kohn theorem~\cite{HohKoh-PR-64} ({\tt cf. chapter \cite{}}).

One of its formulations: ``$\rho$ yields $v_{ne}$ and $N$, and thus everything'' is useless, as we do not need to know the density
 to know the potential of the system under study.
This formulation of the theorem is never used in practice.

However, the variational formulation of the Hohenberg-Kohn theorem is consequential.
It states that
\begin{equation}
\label{eq:hk}
 E [v,w,N]  =  \min_\rho \left(F[\rho,w] +  \displaystyle{\int_{\R^3} \rho(\bfr) v(\bfr) \, d\bfr}  \right) 
\end{equation}
where for $F$ one uses either the Legendre transform form~\cite{Lie-IJQC-83},
\begin{equation}
\label{eq:legendre}
  F[\rho,w]=\sup_v \left( E[v,w,N] - \displaystyle{\int_{\R^3} v(\bfr) \, \rho(\bfr) \, d\bfr} \right)
\end{equation}
or, equivalently, a constrained search for ensembles. 
For the sake of simplicity, in this chapter its pure state form~\cite{Per-78, Lev-PNAS-79, Lie-IJQC-83}
\begin{equation}
  F[\rho,w] = \min_{\Psi \rightarrow \rho}  \bra \Psi T  + W \ket \Psi
 \label{eq:fw}
\end{equation}
 is used.
As $F$ does not depend on $v$ (that specifies the system) the functional is called {\em universal}.
The dependence on $N$ appears through that of $\rho$ (eq. ~\eqref{eq:Nfromrho}).
As above for the energy functional, the dependence on the operator $T$ is not explicited in the notation for $F$.
For the physical system, $w=v_{ee}$, is implicitly assumed; we write:
\begin{equation}
 \label{eq:f}
 F[\rho]=F[\rho,w=v_{ee}]
\end{equation}
The hope raised by eq.~\eqref{eq:hk} is that it can be used with some simple approximation for $F[\rho]$.


\subsection{Difficulty of producing $F[\rho]$}

Obtaining $F$ for a given $\rho(\bfr)$ is possible, but still difficult:
 a constrained minimization, as required by equation~\eqref{eq:fw} is more demanding than 
 a minimization without only the constraint of normalizing the wave function, eq.~\eqref{eq:e0-var}.
The Legendre transformed form of $F$, eq.~\eqref{eq:legendre}, requires first computing $E$ for all $v$, but then no $F[\rho]$ is needed.

Up to now, there is no systematic way to construct approximations for $F[\rho]$.
In practice, $F$ is replaced by some model, $\tilde{F}$: one speaks about a density functional approximation (DFA).

Please notice that using eq.~\eqref{eq:legendre}, due to the variational principle, the errors will be of second order in $v$ for
 first-order errors in $v$. 
Stated differently: there are many $v$ that give values of $F$ that are close.
For example, adding to potential a very rapidly oscillating function yields essentially the same value $F$.
(For this, and other examples, see, e.g., \cite{SavColAll-JCP-01, SavColPol-IJQC-03}.)
Again, it appears to be of little practical importance to follow the line {\em $\rho$ gives $v$ and thus everything}. 
However, obtaining $E_0$ from eq.~\eqref{eq:hk} is not necessarily affected by this problem once $F$ is known or can be approximated.
One can even wonder if the existence of many density functional approximations with similar quality are not due to the indifference
 of $F$ to changes in the approximation of an optimizing $v$ in eq.~\eqref{eq:legendre}.

\section{Practical solutions for density functional approximations}

In order to create models, two main lines have emerged within density functional theory.
\begin{enumerate}
 \item Using a simple ansatz for the density functional.
 \item Considering DFT as an inspiration to develop other methods that do not require an explicit construction of a density functional.
\end{enumerate}
The first approach is  a {\em cutting the Gordian knot} type of solution.
The second approach is in line with methods developed for wave functions, sometimes nothing but such a method.

\subsection{Ansatz}

\subsubsection{Choice of the ansatz}
Most DFAs start with the so-called local density approximation (LDA).
Within this model, a general functional $G[\rho]$ is replaced by the ansatz
\begin{equation}
  \tilde{G}[\rho] = \int_{\R^3} g(\rho(\bfr)) \, d\bfr 
 \label{eq:lda}
\end{equation}
The function $g$ has to be defined in some way. 
Traditionally, it is fixed in the uniform electron gas, a system with an infinite number of particles, 
 and where $\rho$ does not depend on the position {\tt (cf. chapter \cite{})}.
Typically $g$ is either obtained analytically as a function of $\rho$, or computed for a series of values of $\rho$, and fitted to them
 satisfying asymptotic conditions.

LDA has the important advantage of being (to a certain extent) size-consistent, i.e., satisfying
\begin{equation}
 E_{A \dots B} = E_A + E_B
\end{equation}
where $E_{A \dots B}$ is the system composed of two parts, $A$ and $B$, at infinite separation, while $E_A$ and $E_B$ are the energies
of these parts computed independently.
For the violations of size-consistency by LDA, see, e.g.,~\cite{Per-85,Sav-08}.
Another, major, advantage is its computational simplicity (just a numerical integration to obtain $\tilde{G}$), and its linear scaling with system size. 
Both result from the local character of $g$: if $\rho$ can be decomposed into contributions from two spatial parts,
\begin{equation}
  \rho(\bfr) = \{ 
  \begin{matrix}
   \rho_A(\bfr) & \mathrm{for} & \bfr \in \Omega_A(\bfr) \\
   \rho_B(\bfr) & \mathrm{for} & \bfr \in \Omega_B(\bfr)
  \end{matrix}
\end{equation}
so can be $g$; $\tilde{G}$ becomes the sum of the two contributions.

LDA can be extended by making $g$ depend on other local quantities such as derivatives of the density, giving generalized gradient approximations (GGAs),etc.
{\tt (cf. chapter \cite{})}.

\subsubsection{Finding the right functional to approximate by partitioning}
Applying the LDA, eq.~\eqref{eq:lda}, to $F[\rho]$, eq.~\eqref{eq:f}, does not provide the accuracy needed in most electronic structure calculations.
The strategy chosen is to define some density functional $F_d[\rho]$, and approximate only the remaining part, $\bar{F}_d[\rho]=F[\rho]-F_d[\rho]$.

In the following, some choices for the partitioning of $F$ will be presented.
 
\subsubsection{Satisfying electrostatics}
In the classical limit, the electrostatic interaction is given by the nuclear repulsion, 
\[ V_{nn} = \sum_{A,B(>A)} \frac{Z_A Z_B}{\vert \bfR_A - \bfR_B \vert} \]
the interaction between the electron cloud and the positive charges of the nuclei,
\[  \displaystyle{\int v_{ne}(\bfr) \, \rho(\bfr) \, d\bfr} \]
and the repulsion inside the electron clouds, the Hartree energy,
\begin{equation}
  E_H[\rho] = \frac{1}{2} \int_{\R^3} \int_{\R^3} \frac{ \rho(\bfr) \rho(\bfr')}{|\bfr-\bfr'|} \, d\bfr d\bfr'
  \label{eq:eh}
\end{equation}
There is a balance between these contributions. 
For example, between distant neutral atoms these compensate (there is no $\vert \bfR_A - \bfR_B \vert^{-1}$ term in the limit 
 $\vert \bfR_A - \bfR_B \vert^{-1} \rightarrow \infty$).
This balance is destroyed if $E_H$ is approximated, e.g., by using LDA, eq.~\eqref{eq:lda}.
An excess or deficit of repulsion  produces an unphysical repulsion, or attraction of neutral atoms.
Furthermore, even if this balance is enforced by parametrization for a given system, it is not kept for another, even closely related system 
 (see, e.g., ~\cite{SavCol-JMS-00a}).
The solution to this problem was already proposed in the original Hohenberg-Kohn paper~\cite{HohKoh-PR-64}: $E_H$ is treated exactly, and only
 the remaining part approximated.
 
Finding good models for $F[\rho]-E_H[\rho]$ is still an active field of research; 
 there are already approximations that work well for classes of systems, but one does not have yet a universally applicable model.

\subsubsection{Kohn-Sham method: Imposing the Pauli principle}


The Pauli principle is hidden in the wave function used for defining $F[\rho]$, eq.~\eqref{eq:f}.
A way to impose it is to use a model system, with $F[\rho,w \ne v_{ee}]$, where the Pauli principle is imposed, and use approximations
 for the remaining part.
\begin{equation}
\label{eq:eksw}
 E_0 = \min_\Psi  \left( \bra \Psi T  + V_{ne} + W \ket \Psi  + \bar{E}_{Hxc}[\rho_\Psi,w]   \right)
\end{equation}
where the subscript $\Psi$ indicates that $\rho$ is obtained from this wave function, and 
\begin{equation}
 \bar{E}_{Hxc}[\rho,w] = F[\rho,v_{ee}] - F[\rho,w]
 \label{eq:ehxcw}
\end{equation}
This expression is derived using equations~\eqref{eq:e0-var},\eqref{eq:hk},\eqref{eq:fw}.
In general, one takes into account the remark made above about $E_H[\rho]$, and defines
\begin{equation}
 \label{eq:ehw}
 \bar{E}_H[\rho,w] = \frac{1}{2} \int_{\R^3} \int_{\R^3} \rho(\bfr) \rho(\bfr')\left( \frac{1}{|\bfr- \bfr'|} - w(\bfr,\bfr') \right)
                         \, d\bfr d\bfr' 
\end{equation}
The remaining part, 
\begin{equation}
 \label{eq:excw}
 \bar{E}_{xc}[\rho,w] = \bar{E}_{Hxc}[\rho, w] - \bar{E}_H[\rho,w]
\end{equation}
is called {\em exchange-correlation} energy.

With eq.~\eqref{eq:eksw} one is back to an unconstrained variation of a wave function that is chosen to be anti-symmetric, 
 thus satisfying the Pauli principle.

The problem is made simpler by a proper choice of $w$.
For the Kohn-Sham model, one chooses the simplest one, namely $w=0$.

The Kohn-Sham model is usually presented as a modified Schr\"odinger equation that is obtained by the variation of $\Psi$ in
  eq.~\eqref{eq:eksw},
\begin{equation}
 \label{eq:ksw}
 H(w) \Psi(w) = \mathcal{E}(w) \Psi(w)
\end{equation}
where
\begin{eqnarray}
 \label{eq:hw}
 H(w) & = & T + V_{ne} + V_{Hxc}[\rho, w] + W \\
 \label{eq:Vhxc}
 V_{Hxc} & = & \sum_{i=1}^N v_{Hxc}(\bfr_i) \\
 \label{eq:vhxc}
 v_{Hxc} (\bfr, w) & = & \frac{\delta \bar{E}_{Hxc}[\rho,w]}{\delta \rho(r)}
\end{eqnarray}
Please notice that this step (with the extra problem of the existence of the functional derivative) is not needed to obtain $E_0$.
Furthermore, $\mathcal{E}(w)=E[v_{ne}+v_{Hxc},w,N]$, so that
\begin{equation}
 E_0 = \mathcal{E}(w) + \bar{E}_{Hxc}[\rho_0,w] + \displaystyle{\int_{\R^3} \rho_0 \left( v_{ne}(\bfr) - v_{Hxc}(\bfr,w) \right) \, d\bfr}
\end{equation}
where $\rho_0$ is a minimizing density. 
\footnote{For the standard Kohn-Sham model, $\mathcal{E}(w=0)$ is a sum of orbital energies.}

\subsubsection{Using the model wave function}
                       
One can also use the minimizing model wave function, $\Psi(w)$, and choose to approximate the {\em correlation} density functional
\begin{equation}
 \label{eq:ec}
 \bar{E}_c[\rho,w] = F[\rho,w] - \bra{\Psi(w)} T+W \ket{\Psi(w)} 
\end{equation}  

\subsubsection{Problems that remain after splitting $F$}

Separating $F$ into a part defined, $F_d$, and a remainder to be approximated, $\bar{F}_d$, does not necessarily guarantee that an 
 approximation, like that given in eq.~\eqref{eq:lda}, works better.

Separating the Hartree part, $E_H$, analogously to what was done in eq.~\eqref{eq:excw}, removes a problem, but introduces a new one.
Take the limiting case of one-electron systems.
There is no contribution of the interaction between electrons: $E_{Hxc}=0$.
Thus, calculating exactly the Hartree part means that the remaining part has to cancel $E_H$ exactly. 
But obtaining approximations for $-E_H$ is as difficult as obtaining them for $E_H$, and this was considered not to be reachable
 with approximations of LDA-type.
This is known as the {\em self-interaction} problem.

Another (not unrelated) problem is due to degeneracy.
For example, this appears when we consider two parts of the system far apart, and this 
 even in the simplest molecules like H$_2$, or H$_2^+$ when they are stretched (the internuclear distance goes to infinity).
Then, something related to the Einstein-Podolsky-Rosen effect shows up: an infinitesimal perturbation can produce a drastic change in
 the wave function, the density, etc., but not in the energy.
Unfortunately, this gets in conflict with the general philosophy of constructing DFAs that are aimed to produce significant
 changes in the energies for small changes in the density.
 
One could imagine to detect degeneracy.
However, the model systems, in special the mean-field models (such as Kohn-Sham) do not necessarily have ground states presenting
 the same degeneracy as the physical system: while one can present some degeneracy, the other may not.
While the physical wave functions have the symmetry of the Hamiltonian, the model wave function often breaks symmetry to reduce 
 the energy. 
(Well-known is the breaking of spin symmetry showing up when bonds are stretched.)
The opposite can occur, too: Kohn-Sham system can produce degeneracy, while the latter does not show up when Coulomb interaction is present 
 (see, e.g., Fig. 11 in \cite{SavColPol-IJQC-03}).
 
Even more difficult is the case of near-degeneracy, i.e., when a small change in the parameters characterizing $H$ can produce
 degeneracy.
In this case, detecting degeneracy is not a trivial problem, existing in many-body calculations, too.
The standard approach in such situations is giving up using a single Slater determinant as a reference.

\subsubsection{Problems of the model systems}

By construction, the minimizing model $\Psi(w)$ gives an exact ground state density. 
Some other properties can be reproduced, too. 
Trivially, all the expectation values of local one-particle operators, as they need only the density to compute them.
At first surprisingly, the exact ionization potential can be also obtained.
However, this can be easily understood, as it can be related to the asymptotic decay of the density (see, e.g., ~\cite{LevPerSah-PRA-84}, 
 \cite{FouHofHofOst-CMP-09}).

Often, quantities that are not proven to be reproduced exactly by the model system are nevertheless expected to be good approximations.
However, there is the danger of over-stretching this analogy.
For example, it is fashionable to judge DFAs by their ability to reproduce fundamental gaps (differences between the ionization potentials
 and the electron affinities) from differences between orbital energies (of the lowest unoccupied and highest occupied ones).
This is wrong, however~\cite{PerLev-83,ShaSch-83}.
Let us consider, for example, a system with zero electron affinity.
For a neutral system the Kohn-Sham potential, $v_{ne} + v_{Hxc}$, eq.~\eqref{eq:hw}, decays at large distances as $-1/r$ (see, e.g.,\cite{LevPerSah-PRA-84}),
 we know that it supports Rydberg series.
Thus, its gap (ionization potential) is necessarily larger than its first excitation energy.
In fact, accurate Kohn-Sham orbital energy differences give good approximations to excitation energies.
Let us take the He atom as an example~\cite{SavUmrGon-98}.
An extremely accurate Kohn-Sham potential can be obtained from an extremely accurate density.
The Kohn-Sham one-particle Hamiltonian lowest eigenvalue corresponds to the doubly occupied state (1s).
However, higher eigenvalues exist. 
The next eigenvalue (2s) is $\approx 0.75$ hartree about the lowest one.
It can be compared to the excitation energies to the triplet and singlet ($\approx 0.73$, and $0.76$~hartree, respectively).
However, the fundamental gap of the He atom is of $\approx 0.90$~hartree.
(This comparison should not to be confused with potentials produced by DFAs, as LDA for $E_{xc}$ that generates a potential that does not 
 support excited states, and has a ionization potential of $\approx 0.55$~hartree.)
Thus, in general, a DFA that produces an orbital energy difference that reproduces the exact fundamental gap 
 can be expected not to be a good approximation to the exact Kohn-Sham system.

The preceding discussion leads to a slippery ground.
Could it be that the Kohn-Sham approximations are used because they produce convenient mean-field models?
Could it be that (for specific purposes) they may be better than the exact Kohn-Sham system would be?

\subsection{Refining the approximations}

\subsubsection{Spin densities}
The quality of approximations improves considerably when spin densities are introduced, i.e., when the functional $\tilde{G}$ is made to depend not only
on $\rho$, but on its components, the spin-up, $\rho_\uparrow (\bfr)$, and the spin-down $\rho_\downarrow (\bfr)$ densities, 
\[\rho(\bfr) = \rho_\uparrow (\bfr) + \rho_\downarrow (\bfr) \]
Equivalently, one may add to the dependence on $\rho$ that on the spin polarization
\begin{equation}
 \label{eq:zeta}
 \zeta(\bfr)=\frac{\rho_\uparrow(\bfr) -  \rho_\downarrow(\bfr)}{\rho(\bfr)}
\end{equation}

A justification is brought by the fact that the exchange term acts only for electrons of the same spin, and that correlation is not the same
 for a pair of electrons of different spins as for that between two electrons of the same spin (that are kept apart by the Pauli principle).

\begin{figure}[htb]
 \begin{center}
   \includegraphics[width=.7\textwidth]{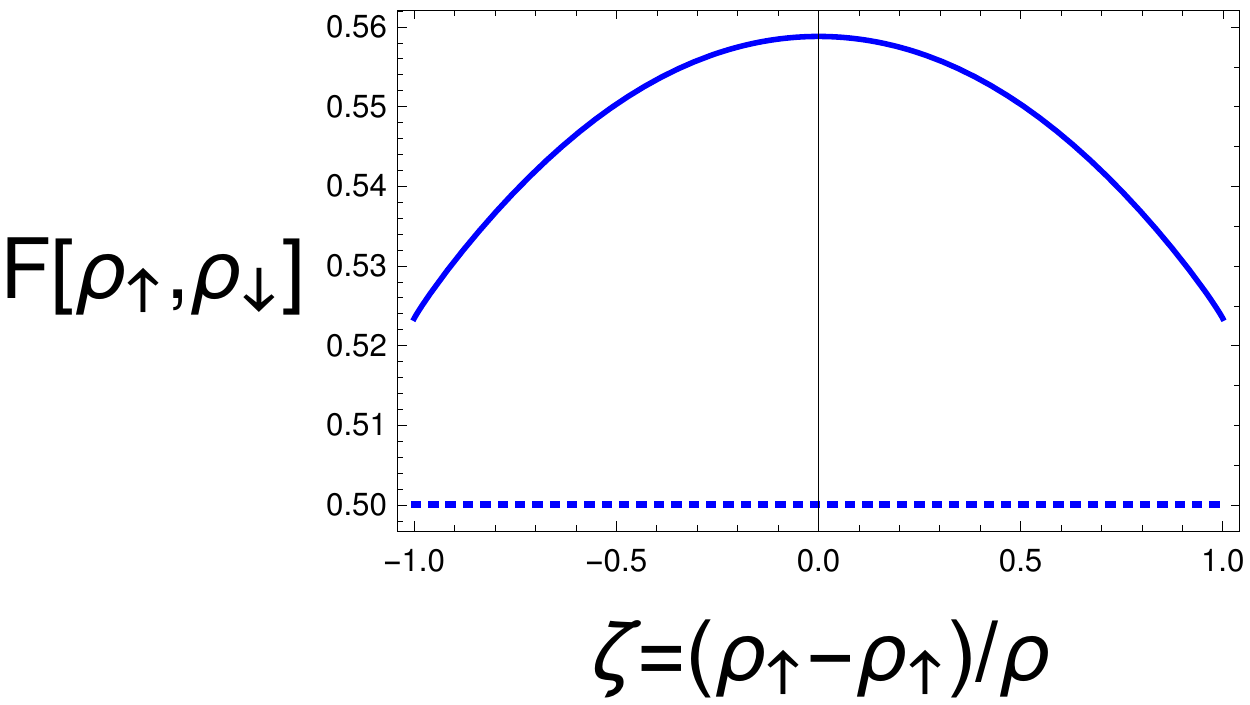}
 \end{center}
 \caption{Dependence of the local density approximation of the functional $F$ on the spin polarization $\zeta$, eq.~\eqref{eq:zeta} 
          for the exact density of the hydrogen atom; 
          the exact value of $F$ is 0.5 hartree (dotted line). }
 \label{fig:finh}
\end{figure}

An example of the importance of making the the functional depend on $\rho_\uparrow$ and $\rho_\downarrow $ is shown in fig.~\ref{fig:finh}.
According to the Hohenberg-Kohn theorem, neither the energy, nor the value of $F$, for the hydrogen atom should depend on $\zeta$.
However, for LDA where a dependence on $\zeta$ is introduced by adjusting the exchange-correlation of the spin-polarized uniform electron gas,
there is a clear dependence on $\zeta$, the best value being obtained when $\zeta=\pm 1$, i.e.,for maximal spin polarization.

\begin{figure}[htb]
 \begin{center}
   \includegraphics[width=.7\textwidth]{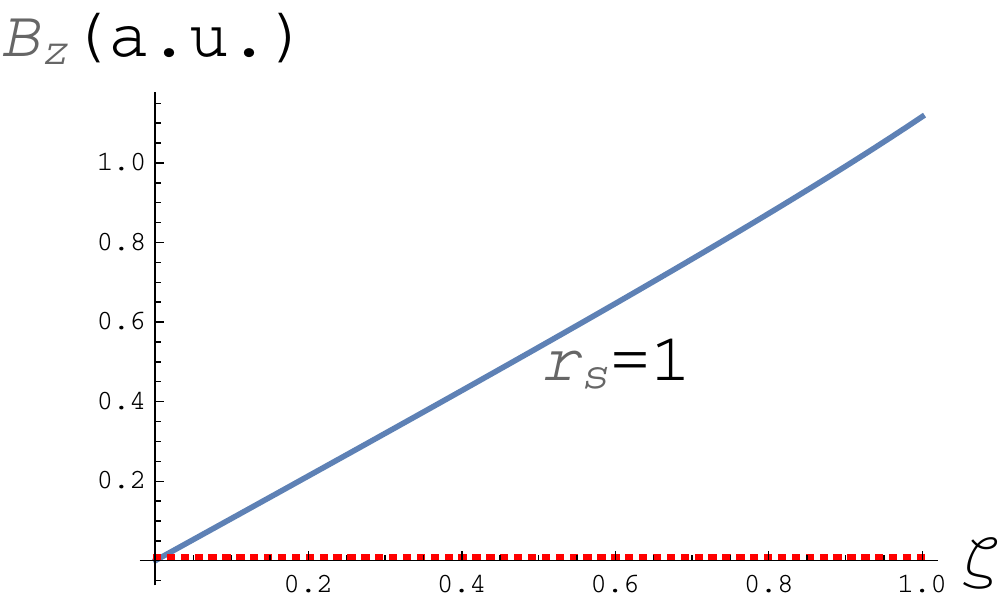}
 \end{center}
 \caption{Strength of the magnetic field, $B$, needed to stabilize the the uniform electron gas with polarization $\zeta$ with
  respect to the unpolarized electron gas with density $\rho=3/4\pi$, i.e., $r_s=1$. 
  The strongest magnetic field ever produced on earth is indicated by a horizontal dotted line.}
 \label{fig:howstrongisb}
\end{figure}

In spite of contributing in a decisive way to the success of DFAs (most achievements in thermo-chemistry would be inexistent 
 without using spin-densities), there is a problem: the theoretical foundation of this approach has never been established.
This affirmation should be supported here by a few of several arguments. 
One hears that the spin-density shows up in a weak magnetic field, and wrongly assumes that 
\begin{enumerate}
 \item a weak magnetic field should not affect the result,
 \item a linear magnetic field should be sufficient, because the field is weak,
 \item it is sufficient to take into account the interaction between the magnetic field and the spins (i.e., only a term $B_z S_z$),
 \item the magnetic fields used for spin-polarized systems are weak.
\end{enumerate}
The first point is wrong, because lifting degeneracy by a magnetic field can produce a different ground state. 
For example, putting the stretched H$_2$ molecule in a weak uniform magnetic field produces a triplet ground state, 
 while in absence of the magnetic field, it is a singlet.
The second point is wrong, because it ignores a general problem:``a small perturbation parameter does not mean a small perturbation''~\cite{Rel-69}.
For the specific case we consider, we notice that even a weak linear magnetic field stabilizes states with high angular momentum
 below the ground state in the absence of the magnetic field.
The variational principle cannot be applied, and the Hohenberg-Kohn theorem cannot be proven~\cite{Sav-MP-17}.
The third point is wrong, as we know from the elementary treatment of the Zeeman effect: the orbital momentum is as important as the spin,
 but if we introduce a dependence on it, we have a dependence on the external potential, and this is not allowed for a universal density functional.
Finally, the forth point is wrong, because in order to produce a spin-polarized electron gas 
 (for densities of chemical interest, $\rho \approx 3/4\pi$, i.e., $r_s=1$) a strong electronic excitation is needed, and this can 
 be produced only by a huge magnetic field (see Fig.~\ref{fig:howstrongisb}).
 
There is, however, a different viewpoint: the spin-density stays for another quantity that can be related to the spin-density.
It has been noticed long ago for unrestricted Hartree-Fock calculations~\cite{YamFue-77} that spin-densities can be connected to on-top pair density,
$P(\bfr,\bfr)$, cf. eq.~\eqref{eq:p}. Starting from
\begin{eqnarray}
 \rho(\bfr)   & = & \rho_\uparrow(\bfr) + \rho_\downarrow(\bfr) \\
 P(\bfr,\bfr) & = & \left( \rho_\uparrow(\bfr) + \rho_\downarrow(\bfr) \right)^2 - \left( \rho_\uparrow(\bfr)^2 + \rho_\downarrow(\bfr)^2 \right)
 \;\; \mathrm{(single \; determinant)}
\end{eqnarray}
an alternative interpretation of the spin-density in DFT is obtained~\cite{BecSavSto-TCA-95,PerSavBur-PRA-95}:
\begin{equation}
 |\rho_\uparrow - \rho_\downarrow| = \sqrt{\rho(\bfr)^2- 2 P(\bfr,\bfr)} \;\; \mathrm{(single \; determinant)}
\label{eq:yama}
\end{equation}
Could it be that the theory behind DFAs is not DFT?

Let us mention that a relationship can be found also between spin-densities and first-order reduced density matrices 
 (\cite{TakFueYam-78}, \cite{StaDav-00}).

\subsubsection{The adiabatic connection}
The adiabatic connection was invoked in order to understand what a density approximation should 
 do~\cite{HarJon-JPF-74, LanPer-SSC-75, GunLun-PRB-76, Yan-98}.
The basic idea is that one constructs a model Hamiltonian depending continuously on a parameter, $H(\mu)$. 
The corresponding Schr\"odinger equation has an eigenvalue $E(\mu)$ and an eigenfunction $\Psi(\mu)$.
We require that for a certain value of this parameter the model Hamiltonian becomes the physical one.
Let us now choose $\mu=\infty$ for it. 
Furthermore, we assume that the Hellmann-Feynman theorem (or first-order perturbation theory) can be applied to this model:
\begin{equation}
\label{eq:helfey}
 \frac{d}{d \mu} E(\mu) = \bra{\Psi(\mu)} \partial_\mu H(\mu) \ket{\Psi(\mu)}
\end{equation}
Suppose that the model system, say at $\mu_0$, is accessible (for example, it is a Kohn-Sham calculation).
We want to know how to correct the model energy, $E(\mu_0)$, to obtain $E=E(\mu=\infty)$.
For the missing part, we use the notation, $\bar{E}(\mu_0)$:
\begin{equation}
\label{eq:ebar}
 E = E(\mu_0) + \bar{E}(\mu_0)
\end{equation}
By integrating eq.~\eqref{eq:helfey} we get what is also called integrated Hellmann-Feynman formula~\cite{EpsHurWyaPar-67}
\begin{equation}
\label{eq:inthelfey}
 \bar{E}(\mu_0) = E - E(\mu_0)  = \int_{\mu_0}^\infty \bra{\Psi(\mu)} \partial_\mu H(\mu) \ket{\Psi(\mu)} \, d\mu
\end{equation}
If we consider (as above) that the model only changes $V$ and $W$, we also write
\begin{equation}
  \label{eq:inthelfeymu}
   E - E(\mu_0)  =  \int_{\mu_0}^\infty \bra{\Psi(\mu)} \partial_\mu \left( V(\mu) + W(\mu) \right) \ket{\Psi(\mu)} \, d\mu
\end{equation}
In density functional theory, one furthermore assumes that one can choose $V(\mu)$ such that the density does not change with $\mu$.
Using eqs.~\eqref{eq:1potential},\eqref{eq:rho}, and the convention used here that $v \rightarrow v_{ne}$ when $\mu \rightarrow\infty$ , 
 we can write
\begin{equation}
  \label{eq:inthelfeydft}
   E - \bra{\Psi(\mu_0)} T + W(\mu_0) +V_{ne} \ket{\Psi(\mu_0)} =  \int_{\mu_0}^\infty \bra{\Psi(\mu)} \partial_\mu  W(\mu)  \ket{\Psi(\mu)} \, d\mu
\end{equation}
Using a relationship analogous to eq.~\eqref{eq:2potential}, and eq.~\eqref{eq:p},
\begin{equation}
 \bra{\Psi(\mu)} \partial_\mu  W(\mu)  \ket{\Psi(\mu)} = \frac{1}{2}\int_{\R^3} \int_{\R^3}  P_\mu(\bfr_1, \bfr_2, \mu) \partial_\mu w(|\bfr_1 - \bfr_2|,\mu) 
   \, d\bfr_1 d\bfr_2 
\end{equation}
Please notice that as $\Psi$ depends on $\mu$, so does $P$.
A comparison with eq.~\eqref{eq:eksw} (where the dependence on $w$ is replaced by that on $\mu$) gives the correction to $E(\mu_0)$:
\begin{eqnarray}
 \bar{E}_{Hxc} (\mu_0)
   & = & \int_{\R^3} d\bfr_1 
   \underbrace{\int_{\mu_0}^\infty d\mu  \int_{\R^3} d \bfr_2 \; P_\mu(\bfr_1, \bfr_2, \mu) \partial_\mu w(|\bfr_1 - \bfr_2|,\mu)}_{e(\bfr_1)}
\label{eq:ehxcad}
\end{eqnarray}
The integrand $e(\bfr_1)$ shows a superficial similarity with the function $g$ appearing in LDA, eq.~\eqref{eq:lda}.
However, unlike LDA, the connection with $\rho$ is not evident.

One can eliminate a known term from $\bar{E}_{Hxc}$, and correct correspondingly the r.h.s.
For example, if we would like to have $\bar{E}_{xc}$, eq.~\eqref{eq:excw}, we eliminate the 
contribution of $\bar{E}_H$, by taking the derivative w.r.t. $\mu$ in eq.~\eqref{eq:ehw}, i.e., by subtracting $\rho(\bfr) \rho(\bfr')$
from $P$ on the r.h.s. of eq.~\eqref{eq:ehxcad}.

\subsubsection{Density or pair-density functional theory?}

Starting from the eq.~\eqref{eq:ehxcad}, one may ask whether one should not construct functionals of the pair density,  $P(\bfr,\bfr')$, instead of one that
 depends on $\rho(\bfr)$.
One can first notice that the pair density, $P(\bfr,\bfr')$, yields, by integration over $\bfr'$, the density $\rho(\bfr)$, up to a factor $N-1$.
The already mentioned relationship between spin-densities and the on-top pair density, eq.~\eqref{eq:yama}, presents itself as a further argument.
However, the conditions to be imposed on $P$ such that it is a fermionic one are difficult, while those to be imposed on $\rho$ are simple 
($\rho$ should be non-negative, and integrate to $N$).

In fact, LDA can be seen as replacing, in each point of space $\bfr_1$, $P(\bfr_1,\bfr_2)$ in eq.~\eqref{eq:ehxcad} by that 
 obtained in the uniform electron gas with density $\rho(\bfr_1)$ (see, e.g., \cite{GunLun-PRB-76}).
This idea can be extended beyond LDA: many successful functionals were constructed starting from this perspective
 (among them those developed by A.D. Becke, or J.P. Perdew and co-workers, see, e.g., \cite{Bec-IJQC-83}).
 
Some people consider the random phase approximation (RPA) as a density functional model.
It can also be seen as constructing a simplified form of $P$ to be used in eq.~\eqref{eq:ehxcad} (see, e.g., \cite{EshBatFur-TCA-12}).

Recently, new approximations using the pair density showed up (see, e.g., \cite{WilVerTruGagCio-17}).

\subsection{Approaching the exact result}

\subsubsection{Limitations of the mean field model}

Even if by miracle we had the exact Kohn-Sham determinant (and potential), we still would miss information about the exact system 
 (with Coulomb interaction).
For example, we still would not have the exact energy.
Unfortunately, the task of obtaining simple functionals capable of dealing with cases when a single Slater determinant is not a 
 good approximation is not solved.
 
Sometimes ensembles of Kohn-Sham states are discussed. 
A formula expressing the correlation energy in terms of weighted Kohn-Sham orbital energies exists~\cite{Sav-95}.
However, we do not know a simple expression for obtaining the weights, and it does not seem that
 they follow a Boltzmann distribution~\cite{SavCol-00}.
 
A long experience in quantum chemistry shows that a single Slater determinant is often a bad starting point for obtaining many properties
 such as the energy. 
There, it seems natural to consider multi-reference methods, i.e., wave functions where more than one determinant deserve a preferential
 treatment.
The selection of determinants is an art, unless selective methods are used, such as CIPSI (configuration
 interaction by perturbation with multiconfigurational zeroth-order wave function selected by iterative process)~\cite{HurMalRan-73}.
In the following, a special way of generating a multi-determinant wave function will be discussed, namely using some (ideally) weak interaction
 operator $W$.
Degenerate (and near-degenerate) states are detected by such operators, and this automatically introduces more than one Slater determinant if
 needed.
Using more complicated wave functions is a price to pay for getting forms that make existing DFAs closer to a theoretically justifiable form.

\subsubsection{Choosing $w$}

Eq.~\eqref{eq:ehxcad} suggests that it may be more easy to obtain approximations for $E_{hxc}$ when $w\ne0$, i.e., $\mu>0$.
Indeed, if $\partial_\mu w$ is short-ranged, we can use some approximation of $P(\bfr,\bfr')$ that is valid only when
$\bfr'$ is close to $\bfr$, and use an expansion around $\bfr$.
In the limit of zero-range ($\delta$-function) we obtain the on-top density $P(\bfr,\bfr)$ that for a single Slater determinant produces
 a connection to the spin-density (eq.~\eqref{eq:yama}), i.e., a form that resembles LDA with spin-dependence.
Furthermore, expanding $P$ in $\bfr'$ around $\bfr$ produces {\em semi-local} terms such as density derivatives~\cite{GilAdaPop-MP-96}.
Finally, we can expect a better transferability between systems when electrons are close, justifying the transferability from other
 systems like the uniform electron gas, in other words, expecting ``universality''.

Also, it seems advantageous to avoid using $w$ that posses a singularity (like the Coulomb interaction), because this induces a strong dependence
 on the basis set used, a very slow convergence to the exact results (cf. the difficulty of converging 
 $\bra{\Psi} \delta(r_{12}) \ket{\Psi}$ with a finite basis set~\cite{Dav-63}).

A simple and computationally convenient form for $w$ satisfying the requirements above is given by 
\begin{equation}
 \label{eq:werf}
 w(\bfr,\mu) = \frac{\mathrm{erf(\mu \vert \bfr \vert)}}{\vert \bfr \vert}
\end{equation}
Its derivative is short-ranged,
\begin{equation}
 \label{eq:dwerf}
 \partial_\mu w(\bfr ,\mu) = \frac{2}{\sqrt{\pi}} e^{-\mu^2 \vert \bfr  \vert^2} 
\end{equation}
and, when $\mu$ is very large
\begin{equation}
\label{eq:dwerfinf}
  \partial_\mu w(\bfr ,\mu) \rightarrow \frac{2 \pi}{\mu^3} \delta(\bfr), 
   \hspace{1cm}  \mathrm{for} \; \mu \rightarrow \infty
\end{equation}
The interaction in eq.~\eqref{eq:werf} that also has the properties:
\begin{itemize}
 \item $w \rightarrow v_{ee}$ when $\mu \rightarrow \infty$,
 \item $w = 0$ when $\mu=0$
\end{itemize}
i.e., by changing $\mu$ it is possible to switch between the Kohn-Sham and the physical system.
This allows considering this method systematically improvable, in the sense that increasing $\mu$ brings the model closer to the 
 physical Hamiltonian.


However, we do not know how far we have to get away from $w=0$ to get reliable approximations.
This can be explored numerically.

\subsubsection{Errors of DFAs for $w>0$}
Below are results obtained with $w$ given by Eq.~\eqref{eq:werf} and the dependence on $\mu$ is analyzed.

First, to construct a density functional approximations to $\bar{E}_{xc}$, Eq.~\eqref{eq:excw}, uniform electron gas calculations are
 used~\cite{Sav-INC-96,PazMorGorBac-PRB-06}.
Now, the LDA, Eq.~\eqref{eq:lda} can be applied to $\tilde{G}=\bar{E}_{xc}$ for any value of $\mu$.

The numerical results given below are for the 2-electron harmonium, a system with the Hamiltonian
\begin{equation}
 \label{eq:harmonium}
 H = T + \sum_{i=1}^2 \omega^2 r_i^2 + \mathrm{erf}(\mu |\bfr_1 - \bfr_2|)/|\bfr_1 - \bfr_2|
\end{equation}
The variables can be separated, and the solutions can be found for real values of $\mu$ and $\omega$ by solving numerically a one-dimensional
 differential equation (see, e.g., \cite{KarCyr-04}).
For $\omega=1/2$, that is chosen below, analytical solutions are known for the non-interacting ($\mu=0$) and the 
 fully interacting ($\mu=\infty$) system.

\begin{figure}[htb]
 \begin{center}
   \includegraphics[width=.7\textwidth]{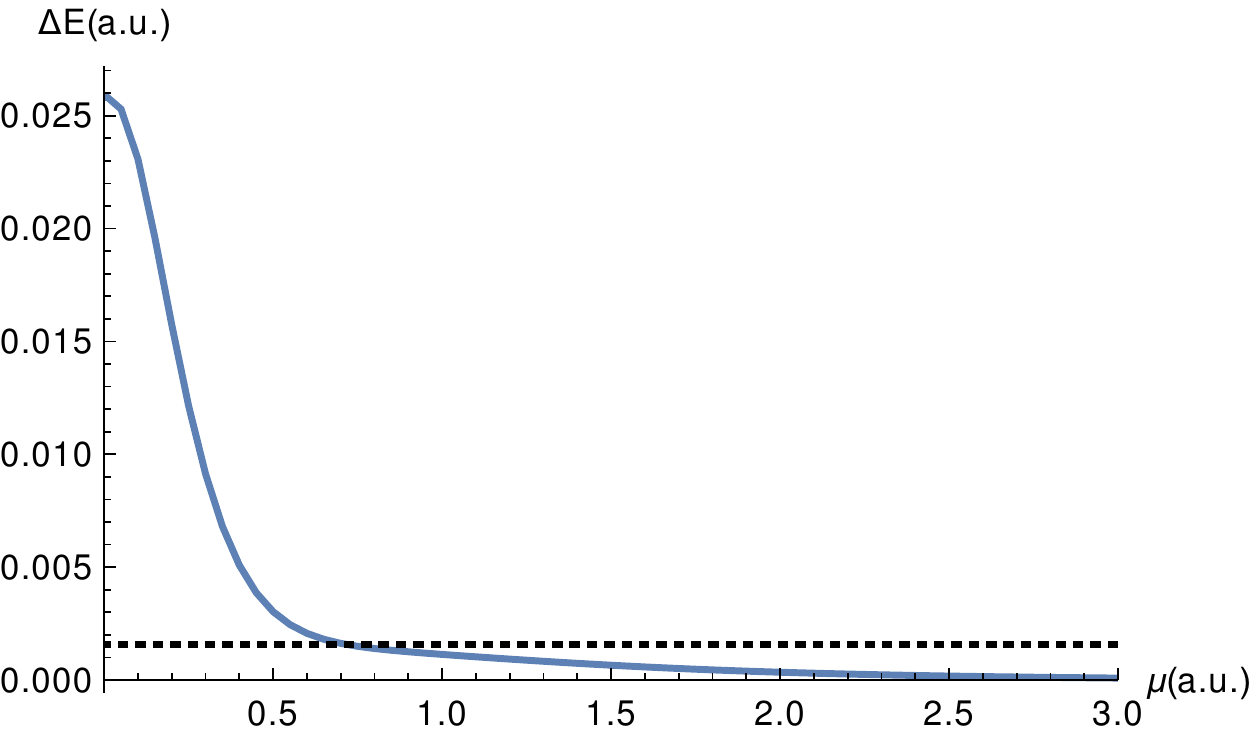}
 \end{center}
 \caption{Errors made by the local density approximation for the ground state energy of harmonium (Eq.~\eqref{eq:harmonium}) as a function of the 
  range separation parameter of the model, $\mu$ (Eq.~\eqref{eq:werf})}
 \label{fig:lda}
\end{figure}

Fig.~\ref{fig:lda} shows the errors made for the harmonium as a function of the choice of the parameter $\mu$.
At $\mu=0$, the error is that given by the usual LDA.
It decreases steadily, and around $\mu=0.5 \dots 1$ a change of behavior occurs, quickly reaching chemical accuracy 
 (1 kcal/mol $\approx$ 2 mhartree).
 
\begin{figure}[htb]
 \begin{center}
   \includegraphics[width=.7\textwidth]{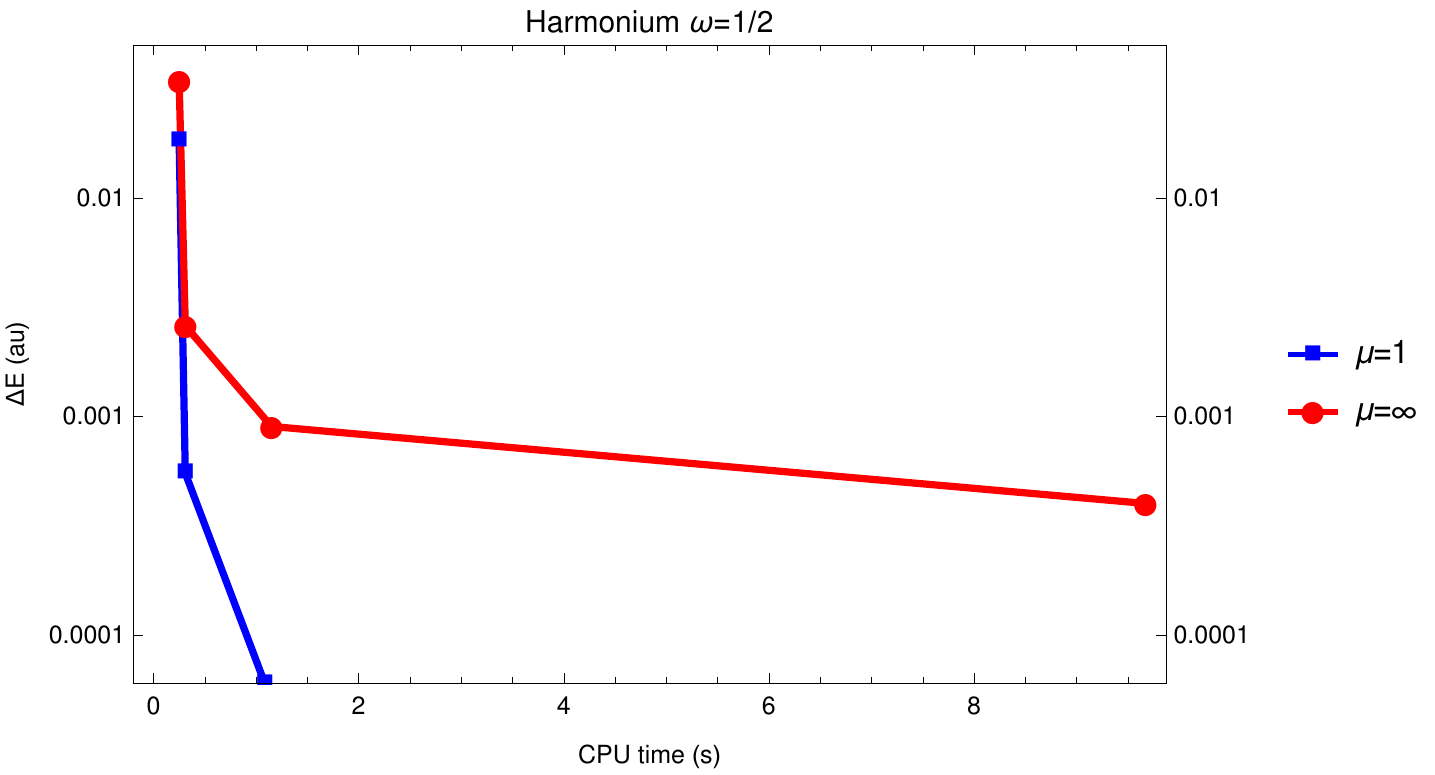}
 \end{center}
 \caption{Harmonium energy errors obtained by saturating the basis set with $l=0,1,2,3$, in a calculation with $\mu=1$, blue, and for
  the Coulomb interaction, red. }
 \label{fig:llimits}
\end{figure}

As having $w \ne 0$ requires having more than a Slater determinant, the time required for computing the wave function rapidly increases
 with $\mu$.
However, as the convergence with the basis set is faster for $w$ having no singularity, the computational effort for obtaining the wave function
 is smaller.
Fig.~\ref{fig:llimits} shows the error that can be achieved in a given time.
Calculations were done first for spherically symmetric basis functions to saturation (s-limit).
Next a new value was obtained for the p-limit ($l=1$), next for the d-limit, ($l=2$), etc.
For such a small system, there is no gain in computing the integrals.
However, one can see that one reaches much faster a high accuracy when $\mu=1$ than with $\mu=\infty$.

For this system, choosing a value of $\mu$ between 0.5 and 1 seems to provide a good compromise between the supplementary effort needed to have 
 $w \ne 0$, and having a good density functional approximation.

\subsubsection{Approaching the exact result analytically}

Instead of using {\em universal} models for $P$ in eq.~\eqref{eq:ehxcad}, one can construct corrections for {\em a given} model Hamiltonian
 energies determined by some $v$ and $w$, $\bar{E}=E[v_{ne},v_{ee},N]-E[v,w,N]$, Eq.~\eqref{eq:ebar}.
The role of the approximation is to correct for the difference between the energy of the exact and that of the model system.
We explore whether standard techniques from numerical analysis could compete withe density functional approximations in estimating these
 corrections.
 
Please notice that as $v \rightarrow v_{ne}$ and $w \rightarrow v_{ee}$, the correction vanishes: $\bar{E} \rightarrow 0$.
One can also try to improve the result by using a set of model Hamiltonians for which obtaining the model energy is simpler than
 finding $E[v_{ne},v_{ee},N]$.

In a density functional context it is tempting to use $v$ as given by some density functional approximation, or even to use the 
 potential that yields the exact density $\rho$ (to show the principle of the procedure). 
Below the simplest expression for the external potential is chosen, $v=v_{ne}$.
Of course, this brings the model system very far from the physical system when the interaction $w$ is weak :
 the errors of the model at $w=0$ are a very important part of the total energy. 
For example, for the harmonium studied above, at $\mu=0$ the error is of 0.5 hartree, as shown in Fig.~\ref{fig:lda}.

First, we analyze how the energy of the model system, $E(\mu)$, approaches that of the Coulomb system, i.e., how $E(\mu)$ approaches $E(\mu=\infty)$.
From the large $\mu$ behavior of the interaction $w$ of eq.~\eqref{eq:werf} and eq.~\eqref{eq:dwerfinf}, we derive  
\begin{equation}
\label{eq:largemue}
 E = E(\mu) + a_{-k} \mu^{-k} + a_{-k-1} \mu^{-k-1} + \dots
\end{equation}
The coefficient $a$ in the equation above is proportional to 
\[ \int_{\R^3} P(\bfr,\bfr,\mu=\infty) \, d\bfr \]
The coefficient $b$ is proportional to $a$, and given by the cusp condition, as $\Psi(\mu)$ has to approach 
 $\Psi=\Psi(\mu=\infty)$ when $\mu$ gets large~\cite{GorSav-PRA-06},
$k$ is equal to $2l+2$, $2l$ being the power of the expansion of $P(\bfr,\bfr+\bfu)$ in $\vert \bfu \vert$ around zero.
In particular, for a pair of singlet coupled electrons (of anti-parallel spin), we have $k=2$ and
\begin{equation}
 \label{eq:largemupsi}
 P(\bfr,\bfr,\mu) = P(\bfr,\bfr,\mu=\infty) (1 + \frac{2}{\sqrt{\pi}}\mu^{-1} + \dots) 
\end{equation}
yielding
\[ \frac{a_{-3}}{a_{-2}} = \frac{4 \sqrt{2}}{3 \sqrt{\pi}} \]

\subsection{Taylor series truncation error}

We consider a Taylor series for large $\mu$.
First, we make a change of variable to $x(\mu)$ such that $x$ monotonously approaches 0 as $\mu \rightarrow \infty$,
\begin{equation}
 \label{eq:taylorinx}
 E(x=0)= E(x)- x E'(x) + \frac{1}{2} x^2 E''(x) + \dots
\end{equation}
Using the chain rule, we go back to the $\mu$ variable,
\begin{equation}
 \label{eq:taylorinmu}
 E(\mu=\infty)= E(\mu)- x(\mu) \frac{E'(\mu)}{x'(\mu)}
      + \frac{1}{2} x(\mu)^2 \left[ E''(\mu) -E'(\mu) x''(\mu)/x'(\mu) \right] \left( x'(\mu) \right)^{-2} + \dots
\end{equation}
Obtaining the first derivative of $E$ with respect to the energy is not expensive, because it does not require computing
 a new wave function.
Of course, the cost increases with higher derivatives.

\begin{figure}[htb]
 \begin{center}
   \includegraphics[width=.7\textwidth]{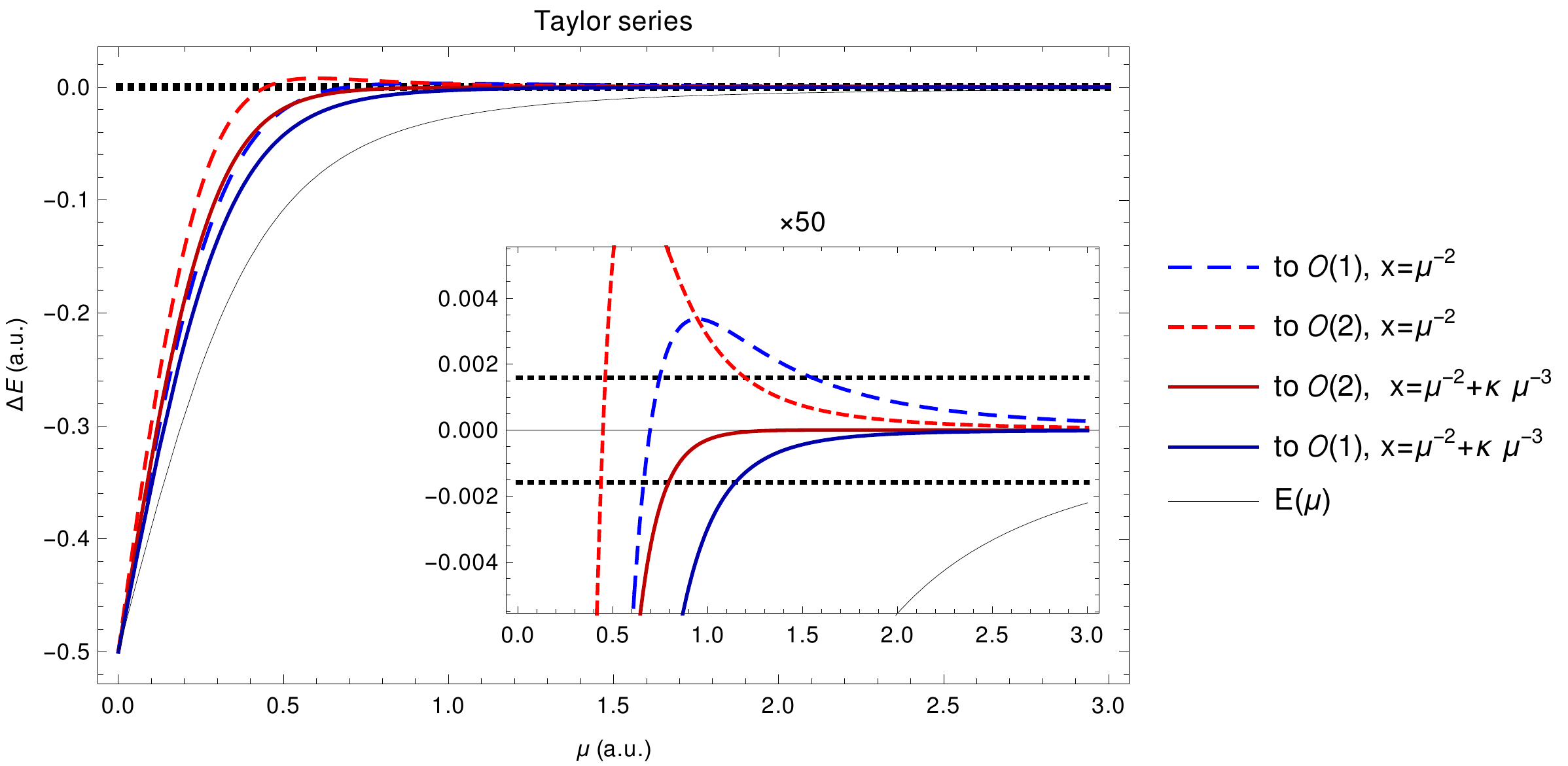}
 \end{center}
 \caption{Errors of different approximations for the ground state energy of harmonium: 
  $E(\mu)$ (thin, black), and the Taylor series around $\mu$ to order 1 (blue), and to order 2 (red);
  the dashed curves correspond to a transformation to $x(\mu)=\mu^{-2}$, the others to $x(\mu)=\mu^{-2}+ \kappa mu^{-3}$.
  The horizontal dotted lines indicate chemical accuracy ($\pm 1$~kcal/mol).
  The inset shows a zoom on the same curves.}
 \label{fig:taylor}
\end{figure}

A choice for change of variable that makes the expansion correct at large $\mu$, is $x=\mu^{-2}$, cf. Eq.~\eqref{eq:largemue}.
As we know the next term in this expansion, we can also choose $x=\mu^{-2} + \kappa \mu^{-3}$.
Not surprisingly, the latter choice is more reliable than the first (cf. Fig.~\ref{fig:taylor}).
However, it is a surprise that the approximation works very well up to $\mu \approx 1$, while the expansion was derived in the limit $\mu \rightarrow \infty$.

Comparing these results with Fig.~\ref{fig:lda}, one notices that the range of models for which the density functional approximation works
 well is comparable to that for which the Taylor series works well.

\section{Outlook}
Most applications of density functional theory rely on the simplicity of using a single Slater determinant. 
This chapter does not intent to discourage the traditional search of density functionals. 
 They are suuccesful in practice, and there still is romm for improvement.
However, using simple mathematical techniques as discussed in the preceding section allows obtaining a good quality, and this is encouraging.
There are many paths that could be followed.
One, of course, is to improve the mathematical techniques.
Another is to change the interaction $w$ to a form for which the extrapolations considered here would work better.
Finally, using density functional models in the connnection withy the extrapolation approach presented here, although this is could be envisaged.

\section{Acknowledgment}
The comments of Eric Canc\`es on the first draft on the manuscript are gratefully acknowledged.

\bibliographystyle{plain}

\begin{thebibliography}{10}

\bibitem{Bec-IJQC-83}
A.~D. Becke.
\newblock {\em Int. J. Quantum Chem.}, {23}:1915, 1983.

\bibitem{BecSavSto-TCA-95}
A.~D. Becke, A.~Savin, and H.~Stoll.
\newblock {\em Theoret. Chim. Acta}, {91}:147, 1995.

\bibitem{Dav-63}
E.~R. Davidson.
\newblock {\em J. Chem. Phys.}, 39:875, 1963.

\bibitem{EpsHurWyaPar-67}
Saul~T. Epstein, Andrew~C. Hurley, Robert~E. Wyatt, and Robert~G. Parr.
\newblock {\em J. Chem. Phys.}, 47:1275, 1967.

\bibitem{EshBatFur-TCA-12}
Henk Eshuis, Jefferson Bates, and Filipp Furche.
\newblock Electron correlation methods based on the random phase approximation.
\newblock {\em Theor. Chem. Acc.}, 131:1084, 2012.

\bibitem{FouHofHofOst-CMP-09}
S.~Fournais, M.~Hoffmann-Ostenof, Th. Hoffmann-Ostenhof, and Th.~Ostergaard
  Sorensen.
\newblock {Analytic Structure of Many-Body Coulombic Wave Functions}.
\newblock {\em Commun. Math. Phys.}, 289:291, 2009.

\bibitem{GilAdaPop-MP-96}
P.~M.~W. Gill, R.~D. Adamson, and J.~A. Pople.
\newblock {\em Mol. Phys.}, 88:1005, 1996.

\bibitem{GorSav-PRA-06}
P.~Gori-Giorgi and A.~Savin.
\newblock {\em Phys. Rev. A}, 73:032506, 2006.

\bibitem{GunLun-PRB-76}
O.~Gunnarsson and B.~I. Lundqvist.
\newblock Exchange and correlation in atoms, molecules, and solids by the
  spin-density-functional formalism.
\newblock {\em Phys. Rev. B}, {13}:4274, 1976.

\bibitem{GutSav-PRA-07}
C.~Gutl\'e and A.~Savin.
\newblock {\em Phys. Rev. A}, 75:032519, 2007.

\bibitem{HarJon-JPF-74}
J.~Harris and R.~O. Jones.
\newblock The surface energy of a bounded electron gas-solid.
\newblock {\em J. Phys. F}, 4:1170--1186, 1974.

\bibitem{HohKoh-PR-64}
P.~Hohenberg and W.~Kohn.
\newblock Inhomogeneous electron gas.
\newblock {\em Phys. Rev.}, {136}:B 864, 1964.

\bibitem{HurMalRan-73}
B.~Huron, J.-P. Malrieu, and P.~Rancurel.
\newblock {\em J. Chem. Phys.}, 58:5745, 1973.

\bibitem{KarCyr-04}
J.~Karwowski and L.~Cyrnek.
\newblock {\em Ann. Phys. (Leipzig)}, 13:181, 2004.

\bibitem{LanPer-SSC-75}
D.~C. Langreth and J.~P. Perdew.
\newblock {\em Solid State Commun.}, 17:1425, 1975.

\bibitem{Lev-PNAS-79}
M.~Levy.
\newblock {\em Proc. Natl. Acad. Sci. U.S.A.}, 76:6062, 1979.

\bibitem{LevPerSah-PRA-84}
M.~Levy, J.~P. Perdew, and V.~Sahni.
\newblock {\em Phys. Rev. A}, 30:2745, 1984.

\bibitem{Lie-IJQC-83}
E.~H. Lieb.
\newblock {\em Int. J. Quantum Chem.}, {24}:24, 1983.

\bibitem{PazMorGorBac-PRB-06}
Simone Paziani, Saverio Moroni, Paola Gori-Giorgi, and Giovanni~B. Bachelet.
\newblock {\em Phys. Rev. B}, 73:155111, 2006.

\bibitem{Per-78}
J.~Percus.
\newblock {\em Int. J. Quantum Chem.}, 13:89, 1978.

\bibitem{Per-85}
J.~P. Perdew.
\newblock In R.~M. Dreizler and J.~da~Providencia, editors, {\em Density
  Functional Methods in Physics}, page 265. Plenum, New York, 1985.

\bibitem{PerLev-83}
J.~P. Perdew and M.~Levy.
\newblock {\em Phys. Rev. Lett.}, 51:1884, 1983.

\bibitem{PerSavBur-PRA-95}
J.~P. Perdew, A.~Savin, and K.~Burke.
\newblock {Escaping the symmetry dilemma through a pair-density interpretation
  of spin-density functional theory}.
\newblock {\em Phys. Rev. A}, 51:4531, 1995.

\bibitem{Rel-69}
F.~Rellich.
\newblock {\em Perturbation Theory of Eigenvalue Problems}.
\newblock Gordon and Breach, New York, 1969.

\bibitem{Sav-95}
A.~Savin.
\newblock {\em Phys. Rev.}, 52:4531, 1995.

\bibitem{Sav-INC-96}
A.~Savin.
\newblock On degeneracy, near degeneracy and density functional theory.
\newblock In J.~M. Seminario, editor, {\em Recent Developments of Modern
  Density Functional Theory}, pages 327--357. Elsevier, Amsterdam, 1996.

\bibitem{Sav-08}
A.~Savin.
\newblock {\em Chem. Phys.}, 356:91, 2009.

\bibitem{Sav-MP-17}
A.~Savin.
\newblock {\em Mol. Phys.}, 115:13, 2017.

\bibitem{SavCol-JMS-00a}
A.~Savin and F.~Colonna.
\newblock {\em J. Mol. Struct. (Theochem)}, {501-502}:39, 2000.

\bibitem{SavCol-00}
A.~Savin and F.~Colonna.
\newblock {\em J. Mol. Struct. (Theochem)}, {527}:121, 2000.

\bibitem{SavColAll-JCP-01}
A.~Savin, F.~Colonna, and M.~Allavena.
\newblock {\em J. Chem. Phys.}, 115:6827, 2001.

\bibitem{SavColPol-IJQC-03}
A.~Savin, F.~Colonna, and R.~Pollet.
\newblock Adiabatic connection approach to density functional theory of
  electronic systems.
\newblock {\em Int. J. Quantum Chem.}, {93}:166, 2003.

\bibitem{SavUmrGon-98}
A.~Savin, C.~J. Umrigar, and X.~Gonze.
\newblock {\em Chem. Phys. Lett.}, 288:391, 1998.

\bibitem{ShaSch-83}
L.~J. Sham and M.~Schl\"uter.
\newblock {\em Phys. Rev. Letters}, 51:1888, 1983.

\bibitem{StaDav-00}
V.N. Staroverov and E.R. Davidson.
\newblock {\em Chem. Phys. Lett.}, 330:161, 2000.

\bibitem{TakFueYam-78}
K.~Takatsuka, T.~Fueno, and K.~Yamaguchi.
\newblock {\em Theor. Chim. Acta}, 48:175, 1978.

\bibitem{WilVerTruGagCio-17}
L.~Wilbraham, P.~Verma, D.G. Truhlar, L.Gagliardi, and I.~Ciofini.
\newblock {\em J. Phys. Chem. Lett.}, 8:2026, 2017.

\bibitem{YamFue-77}
K.~Yamaguchi and T.~Fueno.
\newblock {\em Chem. Phys.}, 19:35, 1977.

\bibitem{Yan-98}
W.~Yang.
\newblock {\em J. Chem. Phys.}, 109:10107, 1998.

\end{thebibliography}

\end{document}